\documentclass[12pt,preprint]{article}
\usepackage{graphicx}
\usepackage{amssymb}
\usepackage{amsmath}
\usepackage{graphics}
\usepackage{epsfig}
\renewcommand{\vec}[1]{{\bf #1}}
\newcommand{\eqb}{\begin{equation}}
\newcommand{\eqe}{\end{equation}}
\newcommand{\dmb}{\begin{displaymath}}
\newcommand{\dme}{\end{displaymath}}
\newcommand{\pd}{\partial}

\newcommand{\eab}{\begin{eqnarray}}
\newcommand{\eae}{\end{eqnarray}}
\newcommand{\ra}{\right\rangle}
\newcommand{\la}{\left\langle}
\newcommand{\e}{\mbox{e}}
\newcommand{\be}{\begin{equation}}
\newcommand{\ee}{\end{equation}}

\begin{document}
\begin{titlepage}
\begin{flushright} 
HD-THEP-07-6\\ 
KA-TP-05-2007
\end{flushright}
\vspace{0.6cm}
\begin{center}
\Large{A model for 
CMB anisotropies on 
large angular scales}\vspace{1.5cm}\\ 
\large{Michal Szopa$\mbox{}^*$ and Ralf Hofmann$\mbox{}^\dagger$}
\end{center}
\vspace{1.5cm} 

\begin{center}
{\em $\mbox{}^*$ Institut f\"ur Theoretische Physik\\ 
Universit\"at Heidelberg\\ 
Philosophenweg 16\\ 
69120 Heidelberg, Germany}
\end{center}
\vspace{1.0cm}
\begin{center}
{\em $\mbox{}^\dagger$ Institut f\"ur Theoretische Physik\\ 
Universit\"at Karlsruhe (TH)\\ 
Kaiserstr. 12\\ 
76131 Karlsruhe, Germany}
\end{center}
\vspace{1.5cm}
\begin{abstract}
We investigate the possibility that a low-temperature and low-frequency 
anomaly in the black-body spectrum, as it emerges 
when enlarging the Standard Model's gauge-group 
factor U(1)$_Y$ to SU(2) (Yang-Mills scale $\sim 10^{-4}$\,eV), 
explains the discrepancy between the Local Group's velocity as 
directly observed and as inferred by assuming a purely kinematic origin of 
the CMB dipole. This discrepancy determines 
the kinetic term for temperature fluctuations in our 
model. The model can be used to predict the low 
multipoles with $l\ge 2$ in the CMB temperature-temperature 
correlation and to reinvestigate the issue of statistical isotropy. 
\end{abstract}
\end{titlepage}

\section{Introduction\label{Intro}}

The physics of photon progation 
enters an exciting epoch in view of the emergence of a number of 
experimental and observational results that are unexplained 
by present theory \cite{PVLAS,WMAP3,knee}. 

The purpose of the present work is to propose a model designed to 
accomodate a discrepancy between the Local Group's velocity as 
directly {\sl observed} by the motion of galaxies 
and as {\sl inferred} by assuming a purely kinematic origin of 
the CMB dipole. The idea is 
that in addition to the kinematic contribution to the 
CMB dipole there exists a dynamic component which 
(in nonrelativistic approximation) is independent 
of the velocity of the observer. The ultimate cause for the 
discrepancy is then attributed to a low-frequency 
and low-temperature\footnote{By low temperature we mean a 
few times $T_{\tiny\mbox{CMB}}=2.73\,$K.} anomaly in black-body spectra as it arises 
when embedding the U(1)$_Y$-factor of the Standard Model into a new 
SU(2) gauge symmetry: SU(2)$_{\tiny\mbox{CMB}}$ \cite{Hofmann20051,Hofmann20052,
GiacosaHofmann2005,SHG20061,SHG20062}. As a consequence of the nonabelian nature of 
SU(2)$_{\tiny\mbox{CMB}}$ the screening mass of the photon is a 
function of momentum, temperature $T$, and Yang-Mills scale $\Lambda_{\tiny\mbox{CMB}}$. The value of the Yang-Mills 
scale $\Lambda_{\tiny\mbox{CMB}}\sim 10^{-4}\,$eV of the latter follows from the observational fact 
that today's photon propagation is not affected by nonabelian fluctuations (thermodynamically decoupled) 
and completely ignores the thermodynamic ground state of this theory 
(no preferred rest frame for photon propagation, see \cite{Hofmann20052}). 
That is, the thermodynamics of SU(2)$_{\tiny\mbox{CMB}}$ is at the 
boundary between the deconfining and preconfining phase 
(supercooled state \cite{GiacosaHofmann2005}) within the 
present cosmological epoch (no screening of photons). 

\begin{figure}
\begin{center}
\leavevmode
\leavevmode
\vspace{6.3cm}
\includegraphics{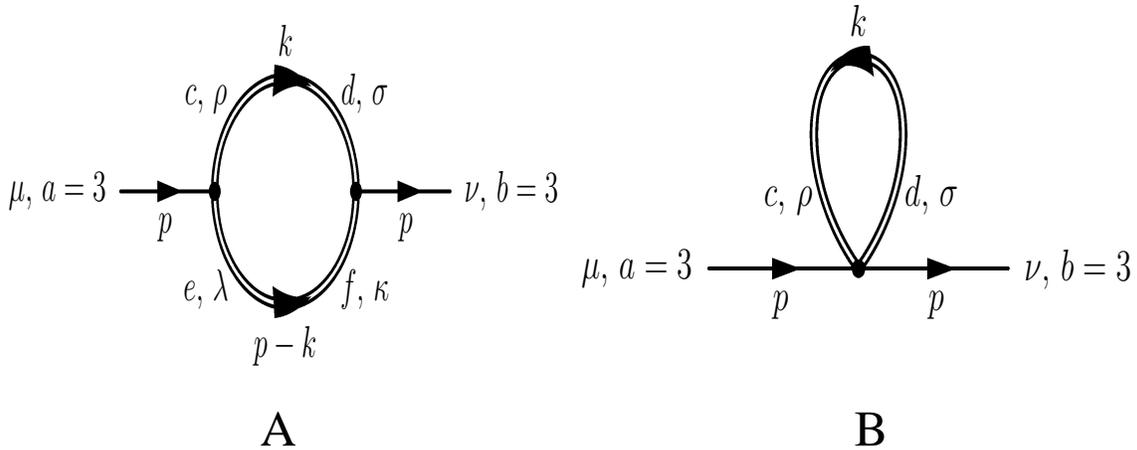}
\end{center}
\caption{\label{Fig-Intro0}The two one-loop diagrams contributing to the polarization tensor
  $\Pi$ of the massless mode (single line). On shell, that is, for
  vanishing external four-momentum squared ($p^2=0$) only the diagram B 
contributes. The local interaction with the
  massive mode (double line) accounts for microscopic photon-monopole
  scattering in a collective fashion.}
\end{figure}

Let us now provide the physical picture of the afore-mentioned
black-body anomaly in terms of the relevant microscopic processes occurring in 
deconfining SU(2) Yang-Mills thermodynamics which seems to underly it. The
thermal ground state in that phase is built up by topologically
nontrivial field configurations: calorons and anticalorons of
topological charge modulus $|Q|=1$. In the hypothetical case of no
interactions between them (anti)calorons essentially exhibit no substructure 
(trivial holonomy). Upon the exchange of trivial-topology gauge-field 
fluctuations (gluons) trivial-holonomy (anti)calorons are 
deformed such that magnetic dipoles\footnote{In terms of the U(1) gauge 
group of electromagnetism the gauge fields and emerging solitons of 
the underlying SU(2) Yang-Mills theory have a dual 
electric-magnetic interpretation: What is an electric field w.r.t. 
the defining SU(2) Lagrangian is a magnetic field w.r.t. U(1) and vice
versa.} emerge which, owing to a radiatively induced attractive
potential \cite{Diakonov2004}, annihilate shortly thereafter. Upon a
spatial coarse-graining down to a consistently prescribed resolution 
$|\phi|$ this situation appears to be spatially homogeneous 
and exerts a negative and temperature dependent pressure. The effect,
which is responsible for the black-body anomaly, however, is not yet 
captured on the level of considering short-lived magnetic dipoles 
only since, after spatial coarse-graining, the diagonal gauge-field 
excitation -- the propagating photon -- 
remains precisely massless. 
Screening or antiscreening of the photon is accounted for by a radiative
correction in the effective theory: The polarization tensor $\Pi$ of the
photon. On shell there seems to be a single one-loop diagram only which
is associated with $\Pi$
\cite{SHG20062,KH2007,Hofmann2006}. Microscopically, diagram B in 
Fig.\,\ref{Fig-Intro0} describes the rarely occurring strong 
deformation of a trivial-holonomy (anti)caloron to yield a magnetic 
dipole whose constituents respulse each other
\cite{Diakonov2004}. 
Monopole and antimonopole now are screened by 
intermediate, short-lived dipoles and thus are isolated, long-lived, and
constitute scattering centers for photon radiation. This is the microscopic 
cause for the black-body anomaly. The effect is maximal for 
temperatures a few times the critical temperature $T_c$. At $T_c$ monopoles
and antimonopoles start to condense into a new ground state, and no more
scattering occurs. This is the present situation. 

For the CMB this means that the radiation released at 
$z=1089$ travels towards us, redshifted by the ever expanding Universe, 
in an almost unadulterated way until the density\footnote{We refer here
  to the relative number density, that is, number of monopoles and 
antimononpoles per volume $T^{-3}$. The absolute density, that is, the
number of monopoles and 
antimononpoles per fixed volume is always increasing with temperature. This
gives rise to a growing spatial string tension and trace anomaly of the
energy-momentum tensor \cite{KortalsAltes,GH2007}.} 
and mobility of isolated monopoles becomes maximal at $z\sim 1$ to leave
an imprint on it. To describe this situation microscopically is, in principle, an
impossible task due to the chaotic motion of these scattering 
centers and the fact that the external probe, needed to resolve this 
motion, would perturb the situation into something far away from
the physical situation in the CMB. The virtue of spatial 
coarse-graining, performed to arrive at the effective theory, 
lies in the fact that scattering effects are correctly accounted
for in a collective and selfconsistent way (no external probes!) 
by a rather simply evaluated one-loop diagram
for the polarization tensor.  

\begin{figure}
\begin{center}
\leavevmode
\leavevmode
\vspace{7.0cm}
\includegraphics{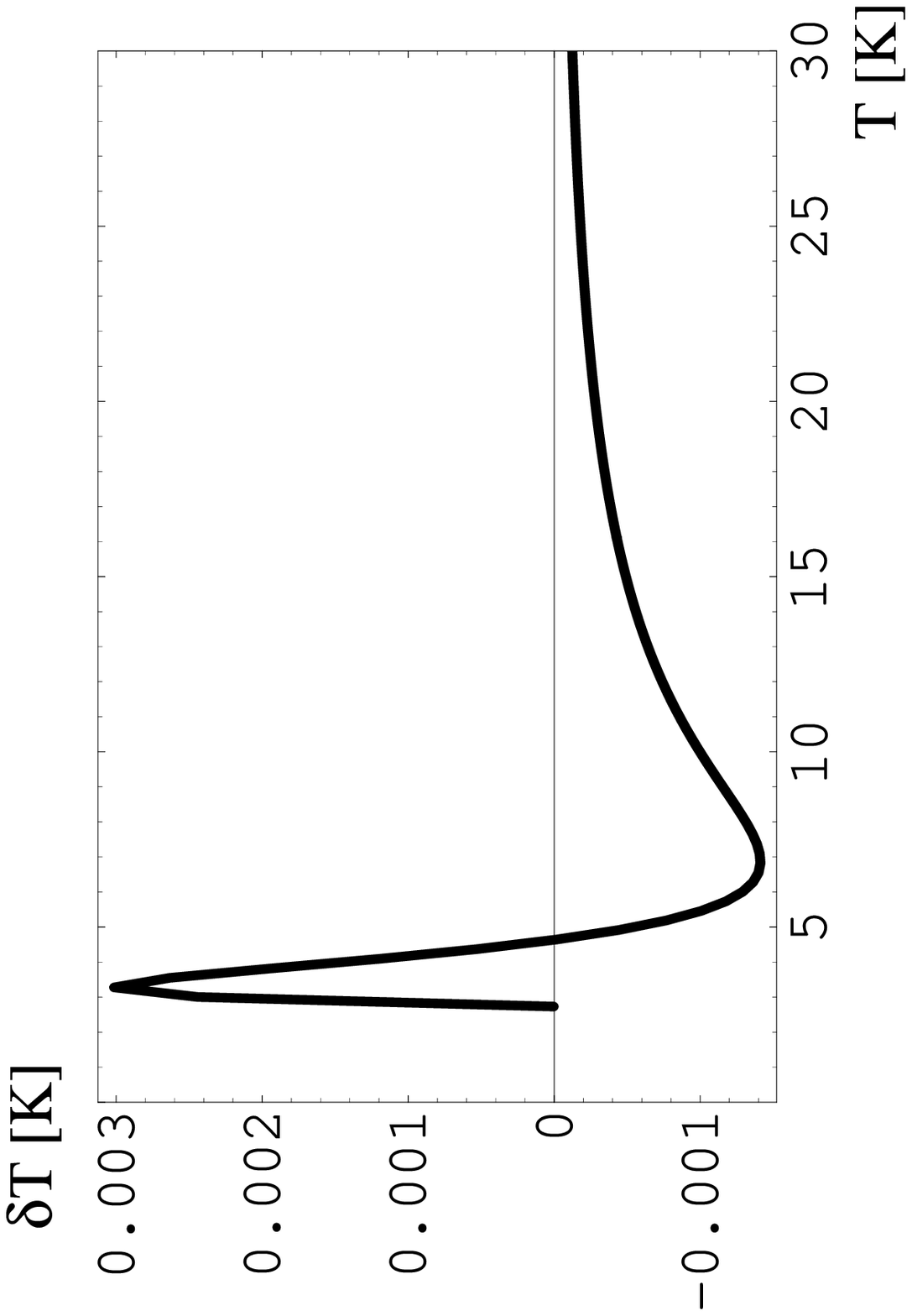}
\end{center}
\caption{\label{Fig-Intro}Temperature offset $\delta T$ as a function of CMB temperature
  $T$. The offset $\delta T$ is extracted from a fit of the black-body
  spectrum, modified by the scattering effects due to the presence of
  isolated monopoles and antimonpoles, to the spectral intensity
  representing the conventional black-body. Plot taken from \cite{SHGS2007}.}
\end{figure}

Because the gradient of the induced temperature offset
(as compared to the conventional case of a U(1) theory for photon
propagation) is also maximal in the vicinity of $z=1$, 
see Fig.\,\ref{Fig-Intro} where this offset is plotted as a function of the
CMB temperature, a maximal temperature inhomogeneity (within our
horizon) 
of primordial origin 
is enforced and with increasing time develops into a 
profile whose spatial slope is responsible for a dynamic 
contribution to the presently observed CMB dipole. 
The only free parameter of the model proposed in the present work is 
determined by the value of 
the discrepancy for the Local Group's velocity. This parameter $k$ is a 
measure for the deformability of the infinite-volume, purely 
thermodynamical situation subject to SU(2)$_{\tiny\mbox{CMB}}$ by
external `forces'. In the case of the CMB these  
forces are on one hand an initial inhomogeneity 
of the temperature distribution induced by primordial causes and on the
other hand the time-dependent background cosmology. The quantity $k$
should, as a matter of principle, be determinable by a linear-response
analysis of the undelying thermodynamics but 
we refrain from performing this analysis in the present work and 
rather appeal to Nature's answer to this question. Once the parameter
$k$ 
is extracted 
from the data the model can be used to predict the dipole-subtracted 
temperature-temperature correlation at small 
angular resolution. We expect that 
the model will postdict the observed suppression of and correlation 
between the low-$l$ multipoles \cite{WMAP3} in the 
temperature-temperature angular power spectrum without the need to 
invoke early reionization. 

The article is organized as follows: 
In Sec.\,\ref{anom}, we review the consequences of the themodynamics of SU(2)$_{\tiny\mbox{CMB}}$ in view of 
an anomaly in black-body spectra at low temperature and frequency as obtained in \cite{SHG20061,SHG20062}. 
We then derive the deviation of the energy density as compared to the conventional case.
In Sec.\,\ref{tevol} we set up our model. Namely, we justify the notion of temperature as a scalar 
field and investigate the dynamics of temperature fluctuations as driven by the black-body anomaly. 
In Sec.\,\ref{numana} we discuss the discrepancy between 
observed and inferred Local-Group velocity and explain how this discrepancy 
is accommodated as a dynamic effect within the realm of SU(2)$_{\tiny\mbox{CMB}}$. 
Subsequently, we perform a numerical analysis of our evolution equation 
when assuming spherical symmetry. Finally, we present and interprete our results. 
In Sec.\,\ref{conclusion} we give a summary and discuss future work.

\section{Anomaly in black-body spectra\label{anom}}

The screening effects on photon ($\gamma$) propagation induced by the charged and massive 
vector excitations $V^\pm$, as they 
emerge by virtue of a nontrivial ground state 
\cite{Hofmann20052}, were computed in 
\cite{SHG20061} by evaluating the polarization 
tensor as a function of temperature $T$ and (on-shell) 
momentum $p$. Again, we point out that evaluating the polarization
tensor in the effective theory corresponds to capturing scattering
effects off of isolated monopoles and antimonopoles
microscopically. Only for temperatures $T$ not far above the critical
temperature $T_c$ and only for a modulus of the spatial $\gamma$ 
momentum much smaller than $T$ is this effect sizable. 
For a detailed discussion of the foundations 
of the nonperturbative approach to SU(2) Yang-Mills thermodynamics in the 
deconfining phase see \cite{Hofmann20051,Hofmann2006,Hofmann2007M}. Only this phase is the relevant 
for CMB physics. 
     
Taking screening effects into account, $\gamma$'s dispersion law modifies as
\eqb
\label{u1dispersion}
\omega^{2}=\mathbf{p}^{2}\ \ \ \ \longrightarrow\ \ \ \ \omega^{2}=\mathbf{p}^{2}+G(\omega,\mathbf{p},T,\Lambda)\,,
\eqe
where $\omega$ is the energy of the $\gamma$-mode and $\mathbf{p}$ its spatial momentum. 
While the screening function $G$ is negative (with a small modulus) for large values of 
$|\mathbf{p}|$ (anti-screening) it is positive 
and of sizable value for small $|\mathbf{p}|$ (strong screening). For details see 
\cite{SHG20062}. At the critical temperature 
$T_{c}\equiv T_{\tiny\mbox{CMB}}=\frac{\lambda_{c}}{2\pi}\Lambda_{\tiny\mbox{CMB}}$ 
($\lambda_{c}=13.87$ \cite{Hofmann20051}), where the $V^\pm$ acquire an infinite mass and 
thus decouple thermodynamically, the propagation of $\gamma$ is entirely 
unscreened ($G\equiv 0$) which is in accord with astrophysical
observation. For thermalized photon propagation at temperatures typically prevailing on earth the effect is 
very weak due to a power suppression of $G$ for $T\gg T_{c}\sim 2.73\,$K
and due to the fact that, decreasing the frequency $\omega$, the
frequency $\omega^*$, where a noticable distortion of the conventional 
Planck spectrum sets in, only increases $\propto T^{1/2}$ whereas the
maximum $\omega_m$ of the distribution is at $\omega_m\sim 2.8\,T$.

\begin{figure}
\begin{center}
\leavevmode
\leavevmode
\vspace{4.5cm}
\includegraphics{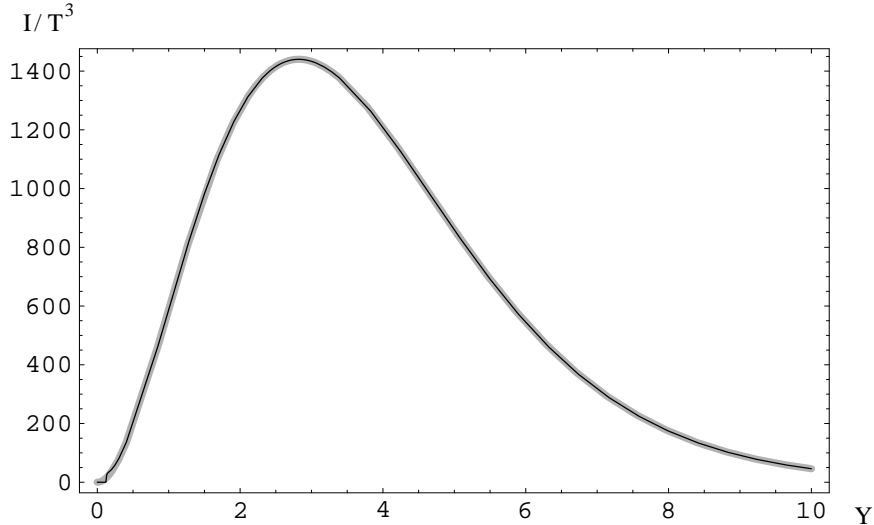}
\end{center}
\caption{Dimensionless spectral power $\frac{I}{T^{3}}$ of a black body as a function of 
dimensionless frequency $Y\equiv\frac{\omega}{T}$ at T=10 K. The black (gray) curve depicts the modified (conventional) spectrum.
\label{Fig-1}}      
\end{figure}

\begin{figure}
\begin{center}
\leavevmode
\leavevmode
\vspace{7.5cm}
\includegraphics{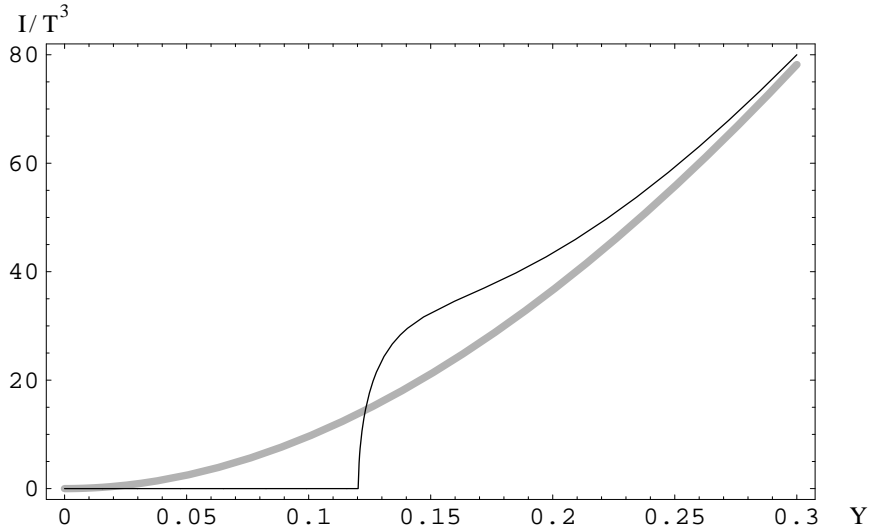}
\end{center}
\caption{Zoom-in of the low-frequency part at T=10 K. The black (gray) curve depicts the modified (conventional) spectrum. 
\label{Fig-2}}      
\end{figure}

According to \cite{SHG20062} the effect of the function \(G\) on the 
spectral power of a black body can be expressed as follows:
\eqb
I_{\mathrm{SU}(2)}(\omega)=I_{\mathrm{U}(1)}(\omega)\times
\frac{(\omega-\frac{1}{2}\frac{\mathrm{d}}{\mathrm{d}\omega}G)\sqrt{\omega^{2}-G}}{\omega^{2}}\theta(\omega-\omega^{*})\,,
\eqe
where $\theta$ is the 
Heaviside step function, $\omega^*$ is the root of $\omega^2=G$, and $I_{\mathrm{U}(1)}$ 
denotes the spectral power of the conventional black body. One has
\eqb
I_{\mathrm{U}(1)}(\omega)=\frac{1}{\pi^{2}}\frac{\omega^{3}}{\exp[\frac{\omega}{T}]-1}\,.
\eqe
Fig.\,\ref{Fig-1} shows the (dimensionless) ratio of the 
modified spectral power $I_{\mathrm{SU}(2)}$ and $T^3$ as a function 
of (dimensionless) frequency $Y\equiv\frac{\omega}{T}$ 
at a temperature of $T=10\,\mathrm{K}$.
In Fig.\,\ref{Fig-2} the low-frequency part of the spectrum, where the deviation from 
the conventional case is best visible, is indicated. 
The deviation $\delta\rho\equiv\rho_{\mathrm{SU}(2)}-\rho_{\mathrm{U}(1)}$ of the energy density then is calculated as
\eqb
\label{deviu1s2}
\delta\rho=\int_{0}^{\infty}\mathrm{d}\omega\,I_{\mathrm{SU}(2)}-\int_{0}^{\infty}\mathrm{d}\omega\,I_{\mathrm{U}(1)}<0\,.
\eqe

\section{Dynamics of temperature evolution\label{tevol}}

\begin{figure}
\begin{center}
\leavevmode
\leavevmode
\vspace{4.5cm}
\includegraphics{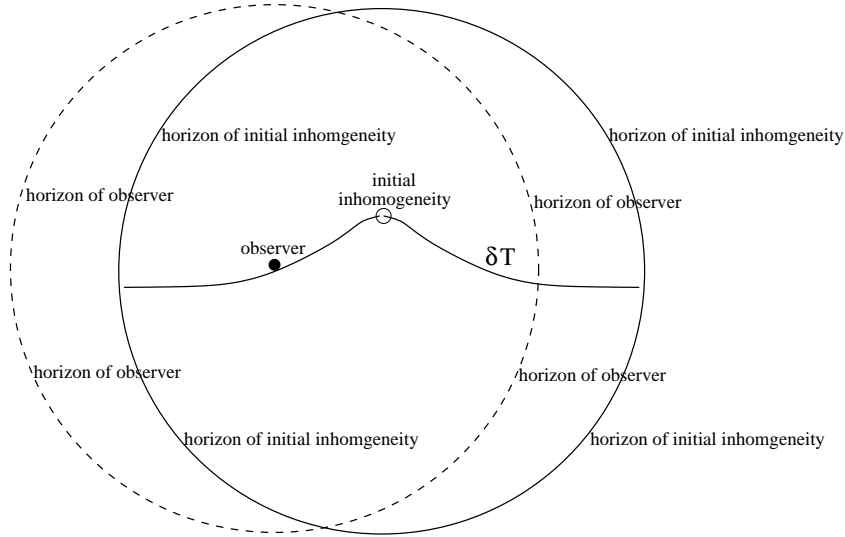}
\end{center}
\caption{The situation relevant for the dynamic contribution to 
the CMB dipole after the black-body anomaly has 
evolved an initial inhomogeneity, for a detailed explanation see text.
\label{Fig-2a}}      
\end{figure}

In this section we derive the dynamic equations 
governing the cosmic evolution of temperature 
fluctuations. In a first step, we perform a match to 
the situation of a perfect fluid which enables us to interprete 
temperature as a scalar field subject to an adiabatic 
approximation. This situation is not unlike the one 
of a thermalized condensed matter system where the evolution of
temperature inhomogeneities is described by a heat 
equation in which temperature plays the role of an scalar field 
under rotations. Subsequently, we allow for deviations from 
the adiabatic limit to obtain a dynamic 
temperature evolution. Finally, we derive the linearized evolution 
equation for temperature fluctuations sourced by the anomaly in the 
black-body spectrum \cite{SHG20061,SHG20062}. This evolution leads 
to the situation as sketched in Fig.\,\ref{Fig-2a}. Namely, an 
initial inhomogeneity, maximal within a sufficiently large spatial 
domain within its horizon, provides for a local gradient of temperature
which, under the influence of the black-body anomaly (scattering of
photons off of monopoles and antimonopoles) is driven to even larger 
gradients as the Universe cools down. Thus a profile 
develops with maximal growth rate in the vicinity of redshift 
$z=1$. As a consequence, an observer situated a certain distance away from the center of
the initial homgeneity measures a dipole distribution of 
CMB temperature within his own horizon. For a terrestial analogue
imagine a volcano which after erruption 2D isotropically 
belches lava across the edge of its crater. A hypothetic 
observer situated at a lower point on the volcano's 
slope preceives a downhill-directed (2D radial) lava flow.        

\subsection{Temperature as a scalar field}

Let us now investigate to what extent it is possible to regard temperature 
as a scalar field. We start be considering the energy-momentum 
tensor $T_{\mu\nu}$ of a perfect fluid whose energy density $\rho$ and pressure $p$ are functions 
of temperature $T$:   
\begin{equation}
\label{perffluidemt}
T_{\mu\nu}=(\rho+p)u_{\mu}u_{\nu}-pg_{\mu\nu}\,,
\end{equation}
where $u_{\mu}$ is the four-velocity of a fluid segment or the local 
rest frame of the heat bath, and the signature of the metric tensor $g_{\mu\nu}$ is 
$(1,-1,-1,-1)$. We now seek an action which produces the right-hand 
side of Eq.\,(\ref{perffluidemt}) upon using the definition for
gravitationally consistent energy-momentum:
\eqb
\label{defT}
T_{\mu\nu}=\frac{1}{\sqrt{-g}}\frac{\delta \mathcal{L}}{\delta g^{\mu\nu}}\,.
\eqe
Here $\mathcal{L}$ is s scalar action density and $\mbox{det}\,g_{\mu\nu}\equiv g$. 
For a static, perfect fluid the only two nonscalar covariants available to
construct this scalar density are $u_\mu$, the local four-velocity of
the fluid, and $g_{\mu\nu}$. Thus we make the ansatz
\begin{equation}
\label{ansatz}
\mathcal{L}=\sqrt{-g}\left(\alpha\,u_{\mu}u_{\nu}g^{\mu\nu}+\beta\right)
\end{equation}
with scalar parameters \(\alpha\) and \(\beta\) to be determined such that the perfect-fluid form in 
Eq.\,(\ref{perffluidemt}) emerges when employing Eq.\,(\ref{defT}). Notice that in varying the action density 
in Eq.\,(\ref{ansatz}) after $g^{\mu\nu}$ the four velocity $u_{\mu}$ is kept fixed. 
The connection between $u_{\mu}$ and $g^{\mu\nu}$ is made subsequently by virtue of Einstein's equations.  

Comparing the coefficients in front of the two independent 
tensor structures in Eq.\,(\ref{perffluidemt}) yields:
\eqb
\alpha=\rho+p\,,\ \ \ \ \ \ \ \ \ \ \beta=-\rho-3p\,.
\eqe
Thus the Lagrangian density in Eq.\,(\ref{ansatz}) becomes 
\begin{equation}
\mathcal{L}=\sqrt{-g}\left((\rho+p)\underbrace{u_{\mu}u_{\nu}g^{\mu\nu}}_{=1}-\rho-3p\right)=-2\sqrt{-g}\,p\,.
\end{equation}
Specializing to a conventional photon gas with equation of state $\rho=3p$, we have
\begin{equation}
\label{idgas}
\mathcal{L}=-\frac{2}{3}\sqrt{-g}\,\rho\,. 
\end{equation}
Since $\rho\propto T^4$ it follows that temperature itself acquires the status of a 
scalar field within the static, perfect-fluid situation. 

\subsection{Temperature fluctuations}

The Lagrangian density in Eq.\,(\ref{idgas}) represents a 
potential for the scalar field $T$. We would like to go beyond this 
adiabatic approximation by allowing for the contributions of derivatives 
in $\delta T$:  
\eqb
\label{Tdec}
T=\bar{T}(t)+\delta T(t,\vec{x})\,.
\eqe
Since $\bar{T}(t)$ is a homogeneous, scalar field corresponding to the 
limit of noninteracting photons the deviation from this limit 
$\delta T(t,\vec{x})$ is also associated with a scalar field\footnote{Notice that the field 
$T$ is a classical field and thus does not admit a particle interpretation of 
its fluctuations.}. 
The deviation $\delta\rho$ of the energy density due 
to the black-body anomaly, see Sec.\,\ref{anom}, induces the fluctuation 
$\delta T$ about the mean temperature $\bar{T}$. The latter is redshifted by the 
evolution of the cosmological background. We consider the conventional black-body part 
$\bar\rho({\bar T})\equiv\rho_{\mathrm{U}(1)}=\frac{\pi^2}{45}\,{\bar T}^4$ 
as a fluid which, among other contributions (cold dark matter and dark energy), sources spatially flat 
Friedmann-Robertson-Walker (background) cosmology\footnote{We are only 
interested in redshifts $z$ up to $z=30$ thus justifying the assumption of a 
spatially flat Universe driven by $\Lambda$CDM.}:
\begin{equation}
\label{FRW}
\mathrm{d}s^2=g_{\mu\nu}\,\mathrm{d}x^{\mu}\mathrm{d}x^{\nu}\equiv \mathrm{d}t^2-a^2(t)\mathrm{d}\vec{x}^2\,,
\end{equation}
where $a(t)$ denotes the scale factor. 

To make the action a scalar the 
usual factor of $\sqrt{-g}$ in the action density 
is expressed in terms of $\bar{T}$ and $\bar{T_0}\equiv T_c$ 
by virtue of the identity
\begin{equation}
\label{redshift}
\frac{a(t)}{a_0}=\frac{\bar T _{0}}{\bar T}\ \ \ \ \Rightarrow \ \ \ \ \sqrt{-g}=\left(\frac{\bar T_0}{\bar T}\right)^3\,,\ \ \ 
(a_{0}\equiv 1)\,.
\end{equation}
In Eq.\,(\ref{redshift}) and in the remainder of the article a subscript `0' refers 
to today's value of the corresponding quantity. Recall that our present
Universe necessarily is very close to $T_c=2.73\,$K to avoid a
contradiction with astrophysical observation: There is no screening
effect in the propagation of photons above the CMB ground state emitted by astrophysical sources 
situated astrophysically (not cosmologically!) far away.   

Leaving the limit of noninteracting photons, the action for $\delta T(t,\vec{x})$ is a 
series involving scalar combinations of arbitrarily high powers of 
derivatives $\pd_\mu$. 
The mass scale $l^{-1}$, which determines the 
relevance of the $n$th power of $\pd_\mu$ in this expansion, is, however, determined by the mass of a 
screened monopole: $l^{-1} \sim \frac{4\pi}{e}\pi T$ where $e=\sqrt{8}\pi$ away from the
deconfining-preconfining phase boundary, recall that there is a
(logarithmic) divergence in $e$ at $T_c$ 
\cite{Hofmann20051,GiacosaHofmann2007}. Thus $l^{-1}$ is comparable 
to temperature itself. Derivatives, however, 
measure deviations of temperature on cosmological scales and thus, for counting purposes, the $n$th 
power of a derivative ($n$ even) can be replaced by the $n$th power of 
the Hubble parameter $H$. For the regime of redshift we are interested in 
($z\le 40$ or so) the parameter $\frac{H}{T}$ is 
extremely small\footnote{In a $\Lambda$CDM model we have $\frac{H}{T}\sim 10^{-33}$ at $z=1$.} 
and thus a truncation of the expansion of the 
action into powers of derivatives at $n=2$ is justified. 
There is not yet {\sl precise} theoretical prescription on how to fix the 
coefficient in front of this kinetic term for $\delta T$ although 
linear-response analysis should be applicable. Being pragmatic, we allow
here for a dimensionless coefficient $k$ whose numerical value 
needs to be determined observationally. 
\noindent Using Eq.\,(\ref{redshift}) we have: 
\begin{equation}
\label{Tdeppa}
\sqrt{-g}\,\mathcal{L}_{\tiny\mbox{CMB}}=\left(\frac{\bar T_0}{\bar T}\right)^3
\left(k\,\pd_\mu\delta T\pd^\mu\delta T-\delta\rho(T)\right)\,.
\end{equation}
Let us now define a function $\hat{\rho}(T,T_0)$ as
\eqb
\label{hatrho}
\delta\rho=T_0^2\,\hat{\rho}\,.
\eqe
Varying the action associated with Eq.\,(\ref{Tdeppa}) w.r.t. $\delta T$
and linearizing the resulting equation of motion then yields:
\eqb
\label{eqmotion}
\partial_{\tilde{\mu}}\partial^{\tilde{\mu}}\delta T-\frac{3}{\bar T}\,\pd_\tau\bar{T}\,\pd_\tau\delta T+
\frac{\bar{T}_0^2}{k H_0^2}\left[\frac12\,\left.\frac{\mathrm{d}^2\hat{\rho}}{\mathrm{d} T^2}\right|_{T=\bar{T}}\,\delta T+\frac12\,
\left.\frac{\mathrm{d}\hat{\rho}}{\mathrm{d} T}\right|_{T=\bar{T}}\right]=0\,.
\eqe
In Eq.\,(\ref{eqmotion}) we have performed the coordinate transformation
\eqb
\label{ct}
\tilde{x}_0\equiv\tau=H_0\,t\,,\ \ \ \ \ \ \ \ \ \tilde{x}_i=\frac{\mathrm{d}a}{\mathrm{d}t}\,x_i\,,\ \ \ \ \ (i=1,2,3)\,.
\eqe
Notice the extremely large factor $(\bar{T}_0/H_0)^2\sim 10^{60}$ in front 
of the square brackets in Eq.\,(\ref{eqmotion}). This factor arises because we chose to 
measure time $\tau$ in units of the age of the Universe, distances from the origin 
$\tilde{x}_i$ in units 
of the actual horizon size $H^{-1}=a/\frac{\mathrm{d}a}{\mathrm{d}t}$ 
(as long as $|\tilde{x}_i|$ is sufficiently smaller than unity), and temperature in units of $\bar{T}_0=2.35\times 10^{-4}\,$eV.   

Assuming spherical symmetry for the fluctuation $\delta T$, which is relevant for an 
analysis of the cosmic dipole, see Sec.\,\ref{BC}, Eq.\,(\ref{eqmotion}) reads:
\eab
\label{difeqO}
0&=&\partial_{\tau}\partial_{\tau}\delta T-\left(\frac{\mathrm{d}a}{\mathrm{d}\tau}\right)^2\,
\left[\partial_{\sigma}\partial_{\sigma}\delta T+\frac{2}{\sigma}\,\partial_{\sigma}\delta T\right]
-\frac{3}{\bar T}\,\pd_\tau\bar{T}\,\pd_\tau\delta T+\nonumber\\ 
&&\frac{\bar{T}_0^2}{k H_0^2}\left[\frac12\,\left.\frac{\mathrm{d}^2\hat{\rho}}{\mathrm{d} T^2}\right|_{T=\bar{T}}\,\delta T+\frac12\,
\left.\frac{\mathrm{d}\hat{\rho}}{\mathrm{d} T}\right|_{T=\bar{T}}\right]\,.
\eae
In Eq.\,(\ref{difeqO}) we have introduced 
$\sigma\equiv\sqrt{\tilde{x}_1^2+\tilde{x}_2^2+\tilde{x}_3^2}$. This
equation is a two-dimen\-sio\-nal wave equa\-tion with
additional terms arising on one hand due to the time-dependence of the cosmological
background ($-\frac{3}{\bar T}\,\pd_\tau\bar{T}\,\pd_\tau\delta T$) and
on the other hand due to the presence of the black-body anomaly: The
term $\frac12
\frac{\bar{T}_0^2}{k
  H_0^2}\left.\frac{\mathrm{d}^2\hat{\rho}}{\mathrm{d}
    T^2}\right|_{T=\bar{T}}\,\delta T$ will be referred to as `restoring term',
and the term $\frac12\,\frac{\bar{T}_0^2}{k
  H_0^2}\left.\frac{\mathrm{d}\hat{\rho}}{\mathrm{d}
    T}\right|_{T=\bar{T}}$ will be referred to as `source term' in the following.    

\subsection{Background evolution\label{DMDE}}

Here we would like to provide some information about the simple $\Lambda$CDM model for the  
background cosmology\footnote{The contribution of $\bar\rho(\bar T)$,
  that is, photon radiation, is 
negligible for $z\le 30$.} which fits the data best \cite{WMAP3,Riess1998,Perlmutter1998,Schmidt1998}. 
We assume a spatially flat Universe subject to the following Friedmann equation
\begin{equation}
\label{Friedmann}
\left(\frac{\dot a}{a}\right)^{2}=
{H_{0}}^{2}\left(\frac{\Omega_{m}}{a^{3}}+\Omega_{\Lambda}\right)
\end{equation}
where \(\Omega_{m}=0.24\) and \(\Omega_{\Lambda}=1-\Omega_{m}=0.76\) (fit obtained from WMAP 
three-year data \cite{Spergel}) are the cold dark-matter 
and the dark-energy density, respectively, both in units of the critical density. 
\(H_{0}\) is today's value of the Hubble parameter, and $\dot{a}\equiv \frac{\mathrm{d}a}{\mathrm{d}t}$. The solution to Eq.\,(\ref{Friedmann}) is
\begin{equation}
\label{solutionback}
a(t)=\left(\frac{\Omega_{m}}{\Omega_{\Lambda}}\right)^{1/3}
\left[\sinh{\frac{3\,\sqrt{\Omega_{\Lambda}}}{2}\,H_{0}t}\right]^{2/3}\,,
\end{equation}
where $H_0$ is connected to \(t_{0}\) (present age of the Universe) as
\eqb
\label{toHo}
H_0 t_0=\frac{1}{3\sqrt{1-\Omega_{m}}}\ln{\frac{2-\Omega_{m}+2\sqrt{1-\Omega_{m}}}{\Omega_{m}}}\,,
\eqe
and we use the convention that \(a_0\equiv a(t_0)=1\). The mean temperature $\bar T$ then follows from 
Eqs.\,(\ref{redshift}) and (\ref{solutionback}). As a reminder we give the relation between scale factor 
$a$ and redshift $z$ since we present our results as functions of $z$:
\eqb
\label{atoz}
a(z)=\frac{1}{1+z}\,,\ \ \ \ \ \ (a_0\equiv a(z=0)=1)\,.
\eqe

\section{Numerical analysis\label{numana}}

\subsection{Principle remarks and boundary conditions\label{BC}}

To identify a dynamic component in the CMB {\sl dipole}, which dominates the higher multipoles 
by two orders of magnitude,  
the associated solution to Eq.\,(\ref{eqmotion}) must locally exhibit a 
singled-out direction. This implies spherical symmetry about the center of an 
initial inhomogeneity which, by the source term in Eq.\,(\ref{difeqO}), 
induces the built-up of the spatially extended, spherically symmetric
profile\footnote{We have, indeed, simulated the full equation (\ref{eqmotion}) not assuming spherical
  symmetry subject to the below-stated boundary conditions. As a result, within
  errors ranging within $\sim$ 1\% the solutions of Eqs.\,(\ref{eqmotion}) 
and (\ref{difeqO}) did coincide. This means that due to the source term
arising from the black-body anomaly a spherical profile with maximum
$\delta T\sim 10^{-2}\,\bar{T}$ is generated and in the process
smoothens out preexisting large-scale fluctuations thus explaining the
latter's suppression.} $\delta T$. 
Notice that a superposition of solutions obtained for several such 
isolated inhomogeneities is not a solution of Eq.\,(\ref{eqmotion}) due to the presence 
of the spatially homogeneous source term. 
Notice also that such an initial situation would evolve\footnote{The dynamic 
situation is possibly not unlike the evolution of an 
initial superposition of static, solitonic configurations in a nonlinear, classical 
field theory as for example the motion of magnetic (anti)monopoles in an SU(2) 
adjoint Higgs model \cite{Atiyah}. Either there is repulsion pushing 
participants beyond each other's horizon or annihilations take place which 
destroy the approximate local spherical symmetry about the center of an 
initial inhomogeneity.} to populate higher multipoles of 
comparable strength as the dipole. This, however, 
is rules out by observation. We conclude that in describing a 
dynamic component to the CMB dipole spherical symmetry of the 
fluctuation $\delta T$ is imperative within the horizon of the center of the 
inducing, initial inhomogeneity. Then \textit{almost} each observer perceives a modulus of the  
dynamic CMB-dipole component which is nearly independent of his position. That is, 
the mean radial gradient approximately serves to 
define a singled-out direction except at the center of the (as we shall see) 
bump-like $\delta T$. This exception, however, 
occurs with vanishing likelihood geometrically, see Fig.\,\ref{Fig-2a}. 

The modulus of the dynamic component $\vec{D}_{\tiny\mbox{dyn}}$, as it 
would be perceived by an observer situated a radial distance 
$\sigma_0$ away from the center of the bump, then is defined as 
follows\footnote{The origin of Eq.\,(\ref{moddyn}) is explained as follows: 
Looking into (opposite to) the direction of the 
gradient, a surplus (deficit) of photons stemming from 
the hotter (colder) tail (central region) of the profile 
$\delta T$ is detected by the observer. This allows 
to define a mean temperature. The amplitude of 
the dipole then is half the difference between the temperature into and opposite to 
the direction of the gradient. These temperatures are 
obtained by performing a radial average over $\delta T$ 
within the horizon of the observer.}
\eqb
\label{moddyn}
|\vec{D}_{\tiny\mbox{dyn}}|\equiv\int^1_{\sigma_0}d\xi\,\delta T(z=0,\xi)-
\int^{\sigma_0}_{\sigma_0-1}d\xi\,\delta T(z=0,\xi)\,.
\eqe
The upper limit in the first integral arises from the fact 
that for $\sigma\ge 1$ the nonexistence of a 
causal connection to the center of the bump forbids the built-up of a 
profile. The definition in Eq.\,(\ref{moddyn}) states 
that $|\vec{D}_{\tiny\mbox{dyn}}|$ is roughly given by 
the mean gradient of $\delta T$. 

Now the coefficient $k$ in Eq.\,(\ref{difeqO}) is determined such 
that the mean gradient in $\delta T(z=0,\sigma)$ coincides with 
the dynamic component in the CMB dipole. The latter is 
attributed to the following discrepancy: On one hand, 
the velocity of the Local Group $\vec{v}_{\tiny\mbox{LG,dir}}$ 
can be determined directly by estimating the gravitational 
impact on it by all those galaxies contained in successively 
enlarged concentric, spherical shells and by observing 
saturation for $z c\ge 6000\,$km\,s$^{-1}$ \cite{2MRS}. 
Here $c$ is the velocity of light. It is found 
that $|\vec{v}_{\tiny\mbox{LG,dir}}|\sim 400\,$km\,s$^{-1}$ with errors 
typically being $\sim 50\,$km\,s$^{-1}$ \cite{2MRS}. 
On the other hand, the conventional understanding of the CMB dipole as 
a purely kinematic effect \cite{Peebles}, which implies a velocity of the solar 
system of $\vec{v}_{\tiny\mbox{SS}}=(369\pm 2)\,
$km\,s$^{-1}$ \cite{Hinshaw}, plus the known relative velocity $\vec{v}_{\tiny\mbox{LG-SS}}$ 
between the solar system and the Local Group allows to deduce a velocity of the Local Group\footnote{In
\cite{PDG2006} a value of $|\vec{v}_{\tiny\mbox{LG,dedu}}|=(627\pm 22)\,$km\,s$^{-1}$ 
was obtained.} of $|\vec{v}_{\tiny\mbox{LG,dedu}}|\sim 619\,$ km\,s$^{-1}$ \cite{Kogut93}. In this way 
the angle $\delta\equiv \angle \vec{v}_{\tiny\mbox{LG,dedu}},\vec{v}_{\tiny\mbox{LG,dir}}$ 
is extracted as $\delta=(13\pm 7)^o$ \cite{2MRS}. 
As a consequence, a deficit velocity 
$\vec{v}_{\tiny\mbox{dyn}}=\vec{v}_{\tiny\mbox{LG,dedu}}-\vec{v}_{\tiny\mbox{LG,dir}}$ is generated 
which must have a {\sl dynamic} origin. 

\begin{figure}
\begin{center}
\leavevmode
\leavevmode
\vspace{6.2cm}
\includegraphics{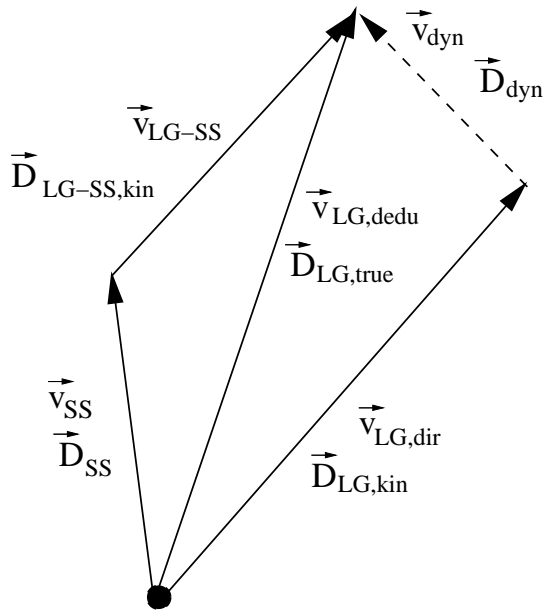}
\end{center}
\caption{Diagram relating observed and deduced quantities concerned 
with CMB-dipole physics.\label{Fig-3}}      
\end{figure}

Let us explain 
this in more detail: On one hand, the velocity $\vec{v}_{\tiny\mbox{LG,dir}}$ generates a kinematic 
contribution to the CMB dipole, $\vec{D}_{\tiny\mbox{LG,kin}}$, whose amplitude 
$\Delta T\equiv\frac12\,|\vec{D}_{\tiny\mbox{LG,kin}}|$ is calculable according to \cite{Peebles} as
\eqb
\label{peebl}
\Delta T=\frac{|\vec{v}|}{c}\,\bar{T}_0+{\cal O}\left(\frac{\vec{v}^2}{c^2}\right)\,.
\eqe
On the other hand, the CMB dipole $\vec{D}_{\tiny\mbox{SS}}$, as it is measured in the solar-system 
rest frame, is supplemented by a purely kinematic contribution $\vec{D}_{\tiny\mbox{LG-SS,kin}}$ 
resulting from the relative velocity $\vec{v}_{\tiny\mbox{LG-SS}}$ between the solar system and the Local Group.
This yields the CMB dipole $\vec{D}_{\tiny\mbox{LG,true}}$ 
as it is perceived in the rest frame of the Local Group. 
Knowing $\vec{v}_{\tiny\mbox{LG,dedu}}$, we can compute 
$\vec{D}_{\tiny\mbox{LG,true}}$ by means of Eq.\,(\ref{peebl}). Since
\eqb
\label{addition}
\vec{D}_{\tiny\mbox{LG,true}}=\vec{D}_{\tiny\mbox{LG,kin}}+\vec{D}_{\tiny\mbox{dyn}}
\eqe
the dynamic contribution to the CMB dipole $\vec{D}_{\tiny\mbox{dyn}}$ follows. 
In Fig.\,\ref{Fig-3} this situation is sketched. 

Let us now discuss the boundary conditions which 
Eq.\,(\ref{difeqO}) needs to be supplemeted with. Eq.\,(\ref{difeqO}) 
is an inhomogeneous, linear, partial differential equation which can be solved using the
numerical method of lines, see \cite{Schiesser}. Four boundary conditions 
are required, two for the temporal and two for the spatial evolution. 
We assume the spatial distribution of the initial fluctuation at redshifts $z_i=5 ... 30$ to be of 
Gaussian shape with its height chosen such that \(\frac{\delta T(z_{i},\sigma=0)}{\bar
T(z_{i})}=10^{-5}\) as is expected to be provided by primordial causes\footnote{We also set
\(\frac{\delta T(z_{i},\sigma=0)}{\bar T(z_{i})}=0\) at times.}: 
\begin{equation}\label{bc1}
\delta T(z_{i},\sigma)=10^{-5}\,\bar T(z_{i})\,\e^{-(\frac{\sigma}{w})^{2}}\,.
\end{equation}
Here the subscript $i$ refers to `initial'. 
The width \(w\) of the Gaussian in Eq.\,(\ref{bc1}) 
will be varied to check for the robustness of the 
result against our ignorance concerning this 
boundary condition. 

\begin{figure}
\begin{center}
\leavevmode
\leavevmode
\vspace{6.3cm}
\includegraphics{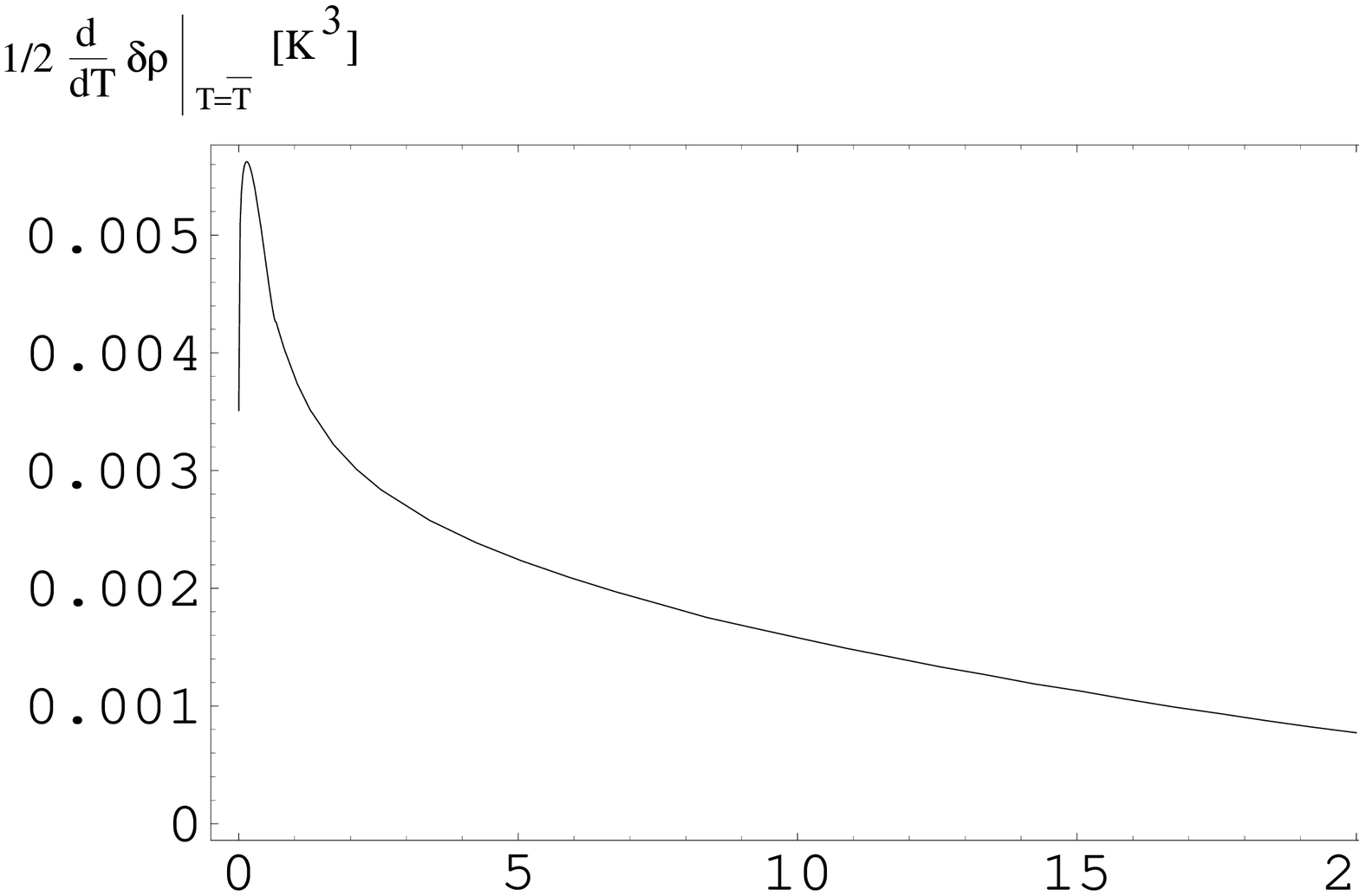}
\end{center}
\caption{The `source term'
  $\frac12\left.\frac{\mathrm{d}\,\delta\rho}{\mathrm{d}
      T}\right|_{T=\bar{T}}$ in Eq.\,(\ref{difeqO}) as a 
function of redshift $z$.\label{Fig-4}}      
\end{figure}

\begin{figure}
\begin{center}
\leavevmode
\leavevmode
\vspace{6.3cm}
\includegraphics{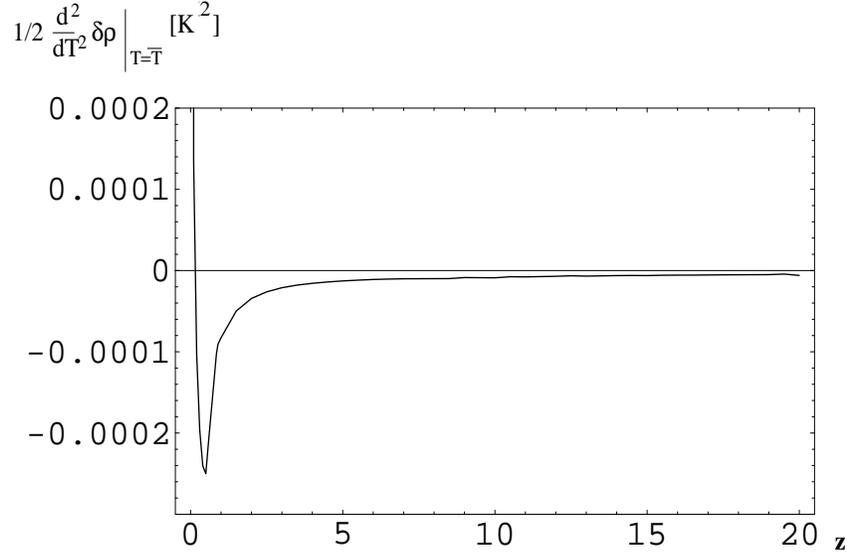}
\end{center}
\caption{The coefficient 
  $\frac12\left.\frac{\mathrm{d^2\delta\rho}}{\mathrm{d}T^2}\right|_{T=\bar{T}}$ in the 
`restoring term' of Eq.\,(\ref{difeqO}) as a function of redshift $z$.\label{Fig-4a}}      
\end{figure}

Initially, we assume the built-up of the fluctuation to be slow since 
the source term $\frac12\left.\frac{\mathrm{d}\,\delta\rho}{\mathrm{d} T}\right|_{T=\bar{T}}$ 
in Eq.\,(\ref{difeqO}) driving this built-up is small for sufficiently large 
initial redshift $z_i$, see Fig.\,\ref{Fig-4}. That is, we prescribe
\begin{equation}\label{bc2}
\partial_{\tau}\delta T(\tau,\sigma)\bigr|_{\tau=\tau_{i}}=0
\end{equation}
and later check for the independence of the result on our choice of $z_i$. 
For a comparison of orders of magnitude 
between the two terms in Eq.\,(\ref{difeqO}) dependent on the black-body anomaly, 
the coefficient
$\frac12\left.\frac{\mathrm{d^2\delta\rho}}{\mathrm{d}T^2}\right|_{T=\bar{T}}$ 
of $\delta T$ in the `restoring term'  
is depicted as a function of $z$ in Fig.\,\ref{Fig-4a}, and in 
Fig.\,\ref{Fig-4} the corresponding plot for the `source term' is
shown. Since our simulations yield $\delta T<10^{-2}\,$K 
we conclude that the `source term' strongly dominates 
the `restoring term'. 
As a function of $\sigma$ 
the fluctuation $\delta T$, being either 
weakened or enforced during the evolution, remains 
extremal at $\sigma=0$ where the initial inhomogeneity (a seed for $\delta T$) 
was located. So it should satisfy 
\begin{equation}\label{bc3}
\partial_{\sigma}\delta T(\tau,\sigma)\bigr|_{\sigma=0}=0\,.
\end{equation}
Finally, $\delta T$ is zero for \(\sigma\ge 1\) 
(horizon\footnote{This statement 
is only approximately valid because the employed relation between coordinates $\tilde{x}_i$ 
and ${x}_i$ in Eq.\,(\ref{ct}) actually is only valid for their differentials.}) and for all times. 
Otherwise, the built-up of $\delta T$ would be noncausal:
\begin{equation}\label{bc4}
\delta T(\tau,\sigma\ge 1)=0\,.
\end{equation}
In order to be consistent 
with the b.c. in Eq.\,(\ref{bc1}), we approximate the b.c. of Eq.\,(\ref{bc4}) by simply 
prescribing the value of the profile $\delta T$ at $z_i=0$ and 
$\sigma=1$ for all $z$. This is in good agreement with Eq.\,(\ref{bc4}) 
if $w\ll 1$. 

\subsection{Values for $k$ and higher multipoles}

The coefficient $k$ in 
Eq.\,(\ref{difeqO}) is chosen such that half of the mean gradient of the profile 
$\delta T$ at $z=0$ equals the observationally inferred amplitude $\Delta T$ for the 
dynamic contribution $|\vec{D}_{\tiny\mbox{dyn}}|$. We impose one observationally suggested set 
($A$) and one set with a fictitiously large angle $\delta$ and an upper-limit 
value for $|\vec{v}_{\tiny\mbox{LG,dir}}|$ ($B$) 
(taking $|\vec{v}_{\tiny\mbox{LG,dedu}}|\sim 627\,$ km\,s$^{-1}$) as
\eab
\label{oba}
(A):&& \ |\vec{v}_{\tiny\mbox{LG,dir}}|=400\,\mbox{km\,s}^{-1}\,,\ \ \ \delta=13^o\ \ \ \Rightarrow\nonumber\\ 
&&\ |\vec{v}_{\tiny\mbox{dyn}}|=253.74\,\mbox{km\,s}^{-1}\,,\ \ \ |\vec{D}_{\tiny\mbox{dyn}}|=2.311\,\mbox{mK}\ \ \ \Rightarrow\ \ \ 
k=0.01868\,\bar{T}_0^2/H_0^2\,;\nonumber\\ 
(B):&& \ |\vec{v}_{\tiny\mbox{LG,dir}}|=450\,\mbox{km\,s}^{-1}\,,\ \ \ \delta=30^o\ \ \ \Rightarrow\nonumber\\ 
&&\ |\vec{v}_{\tiny\mbox{dyn}}|=327.00\,\mbox{km\,s}^{-1}\,,\ \ \ |\vec{D}_{\tiny\mbox{dyn}}|=2.978\,\mbox{mK}\ \ \ \Rightarrow\ \ \ 
k=0.01449,\bar{T}_0^2/H_0^2\,.\nonumber\\ 
\eae
In writing the two values for $k$ in Eq.\,(\ref{oba}) we have anticipated some
results of Sec.\,\ref{numres}. Namely, 
our simulations indicate that the mean gradient of the 
profile $\delta T$ at $z=0$ does not depend on $z_i$ for 
$5\le z_i\le 30$, does not depend on the 
width $w$ in Eq.\,(\ref{bc1}), and does not depend 
on the height for 
\eqb
0\le\delta T(z=0,\sigma=0)\le 10^{-5}\,\bar{T}(z=0)\,.
\eqe 
The mean gradient does, however, depend roughly linearily on the strength of the source term in 
Eq.\,(\ref{difeqO}) thus generating a definite value\footnote{The dependence 
of $|\vec{D}_{\tiny\mbox{dyn}}|$ on $\sigma_0$, as dictated 
by Eq.\,(\ref{moddyn}), is weak in the vicinity of the maximum 
at $\sigma_0\sim\frac23$. Strictly speaking, the observationally inferred value of 
$|\vec{D}_{\tiny\mbox{dyn}}|$ only fixes a curve ${\cal C}$ in the $k$-$\sigma_0$ 
plane, and it must be checked to what 
extent the postdiction of the dipole-subtracted 
correlator depends a variations of $k$ and $\sigma_0$ along ${\cal C}$.} for $k$.    

On one hand, a virtue of the model is to accommodate the possiblity 
for $\vec{D}_{\tiny\mbox{dyn}}$, the latter serving to fix the 
value of the coefficient $k$. On the other hand, once $k$ is fixed by this observational 
input a calculation of the dipole-subtracted large-angle correlation function (with a slight abuse of notation)
\eqb
\label{cordef}
C(\theta)\equiv\la \delta T_{\tiny\mbox{dyn},l\ge 2}(\hat{e}_1),\delta T_{\tiny\mbox{dyn},l\ge 2}(\hat{e}_2)\ra\,,\ \ \ \ \ (\theta\equiv\angle\hat{e}_1,\hat{e}_2)\,,
\eqe
is enabled by virtue of Eq.\,(\ref{eqmotion}). 
Namely, by subtracting Eq.\,(\ref{difeqO}) (dynamic contribution $\delta T_{\tiny\mbox{dyn},l=1}$ to the 
dipole) from Eq.\,(\ref{eqmotion}) 
(general fluctuation $\delta T_{\tiny\mbox{dyn}}$, not imposing 
spherical symmetry) we arrive at
\eqb
\label{highermult}
\partial_{\tilde{\mu}}\partial_{\tilde{\mu}}\delta T_{\tiny\mbox{dyn},l\ge 2}-\frac{3}{\bar T}\,\pd_\tau\bar{T}\,\pd_\tau
\delta T_{\tiny\mbox{dyn},l\ge 2}+
\frac12\,\frac{\bar{T}_0^2}{k H_0^2}\,
\left.\frac{\mathrm{d}^2\hat{\rho}}{\mathrm{d} T^2}\right|_{T=\bar{T}}\,\delta T_{\tiny\mbox{dyn},l\ge 2}=0\,,
\eqe
where $\delta T_{\tiny\mbox{dyn},l\ge 2}\equiv\delta T_{\tiny\mbox{dyn}}-\delta T_{\tiny\mbox{dyn},l=1}$. 
That is, dipole-subtracted fluctuations obey a wave equation 
with a cosmological damping term and a `restoring term' (second 
derivative of the black-body anomaly 
$\delta\rho$ times $\delta T_{\tiny\mbox{dyn},l\ge 2}$). Actually,
Eq.\,(\ref{highermult}) is an approximation assuming that the built-up
of the dynamic contribution to the dipole consumes the entire 
source term and that no influence of this term on 
the higher multipoles takes place. To check whether this, indeed, is the 
case the solution to Eq.\,(\ref{eqmotion}) would have to be projected
onto its multipoles, and the dipole component would have to be 
compared with the solution to Eq.\,(\ref{difeqO}). 
In any case, Cartesian two-point correlations of $\delta T_{\tiny\mbox{dyn},l\ge 2}$ 
at $z=0$, which are required to postdict $C(\theta)$, 
can be computed by an average with primordially provided 
initial conditions of the product 
\eqb
\label{prod}
\delta T_{\tiny\mbox{dyn},l\ge 2}(z=0,\vec{\tilde{x}_1})\,\delta T_{\tiny\mbox{dyn},l\ge 2}(z=0,\vec{\tilde{x}_2})\,,
\eqe
where $\delta T_{\tiny\mbox{dyn},l\ge 2}$ either is 
a solution to Eq.\,(\ref{highermult}) or the according projection onto
the multipole $l\ge 2$ of a solution to Eq.\,(\ref{eqmotion}) 
(classical approximation, see \cite{Tkatchev,Prokopec}). 
The correlation function $C(\theta)$ then follows as
\eqb
\label{corrang}
C(\theta)=\int_0^1 d|\vec{\tilde{x}_1}|\,\int_0^1 d|\vec{\tilde{x}_2}|\,
\la\delta T_{\tiny\mbox{dyn},l\ge 2}(z=0,|\vec{\tilde{x}_1}|\hat{e}_1)\,
\delta T_{\tiny\mbox{dyn},l\ge 2}(z=0,|\vec{\tilde{x}_2}|\hat{e}_2)\ra\,,
\eqe
and one can check the usual assumption made 
about statistical isotropy\footnote{This assumption is heavily contested in 
\cite{Copi2007} based on a large-angle analysis of the WMAP 
three-year data.} by varying $\hat{e}_1, \hat{e}_2$ while keeping 
$\theta$ fixed. This analysis is reserved for future work.

\subsection{Results of numerical calculation\label{numres}}

\begin{figure}
\begin{center}
\leavevmode
\leavevmode
\vspace{4.5cm}
\includegraphics{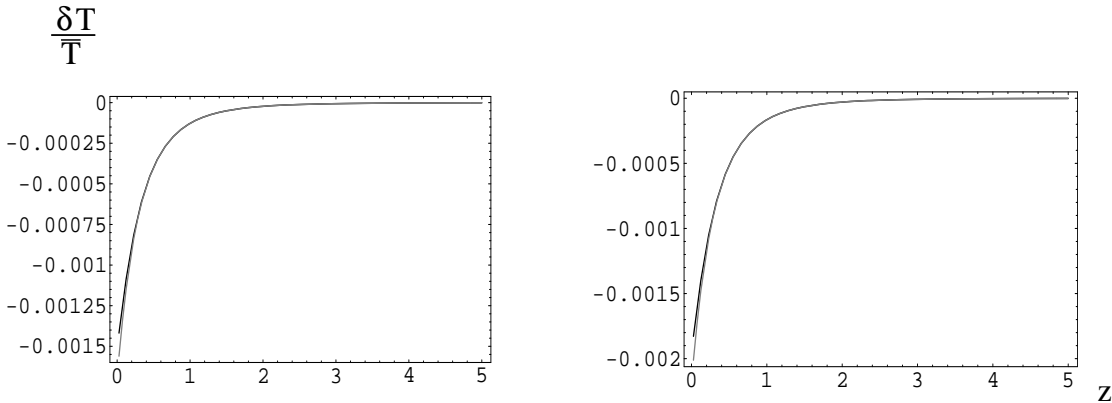}
\end{center}
\caption{$\frac{\delta T}{\bar T}$ at two distances 
$\sigma=0.5$ (black curve) and  $\sigma=0.05$ (gray curve) as a function 
of $z$ setting $z_i=20$. The left (right) panel corresponds to $k=0.01868\,\bar{T}_0^2/H_0^2$ ($k=0.01449\,\bar{T}_0^2/H_0^2$). 
A width $w=10^{-2}$ of the initial Gaussian was assumed.\label{Fig-5}}      
\end{figure}

Here we present the results of our numerical calculation. 
Fig.\,\ref{Fig-5} shows the fluctuation $\delta T$ for $\sigma=0.5; 0.05$ 
as a function of redshift $z$ with a width of the initial Gaussian assumed as $w=10^{-2}$.
Obviously, the major contribution to $\frac{\delta T}{\bar T}$ is generated within $0\le z\le 1$ corresponding 
to a temperature range $T_0=2.73\,\mbox{K}\le T\le 8.1\,$K. This is expected from 
the discussion in \cite{SHG20061,SHG20062}. We have checked that, 
switching off the source term in Eq.\,(\ref{difeqO}) and keeping all 
other conditions the same, the initial profile oscillates in a strongly damped way. 
This is consistent with our finding that the evolution in $z$ as described by 
Eq.\,(\ref{difeqO}) possesses an attractor which is determined by this 
source term, see below. 

\begin{figure}
\begin{center}
\leavevmode
\leavevmode
\vspace{4.5cm}
\includegraphics{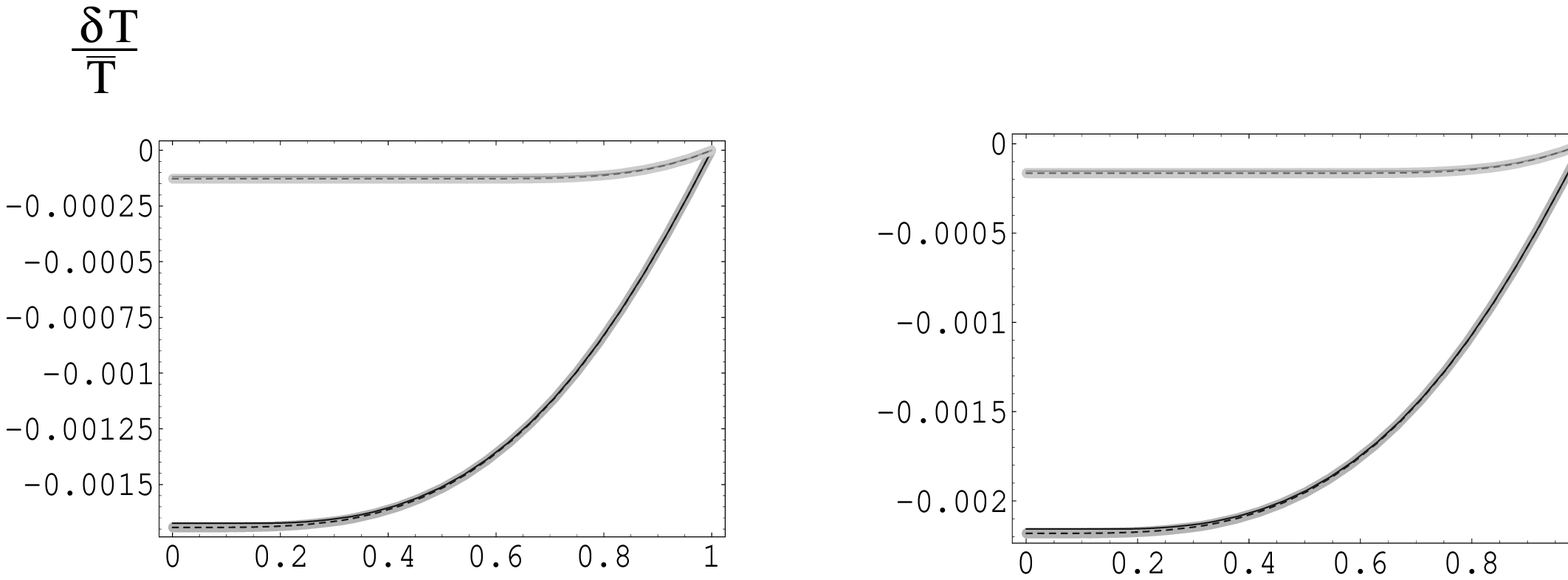}
\end{center}
\caption{$\frac{\delta T}{\bar T}$ as a function of $\sigma$ for a 
width of the initial Gaussian distribution of $w=10^{-2}$ and 
for $z=0$ (lower curves) and $z=1$ (upper curves). The initial redshifts are 
chosen as $z_i=5$, $z_i=20$, and $z_i=40$. 
The left (right) panel corresponds to $k=0.01868\,\bar{T}_0^2/H_0^2$ ($k=0.01449\,\bar{T}_0^2/H_0^2$).
\label{Fig-8}}      
\end{figure}

\begin{figure}
\begin{center}
\leavevmode
\leavevmode
\vspace{4.5cm}
\includegraphics{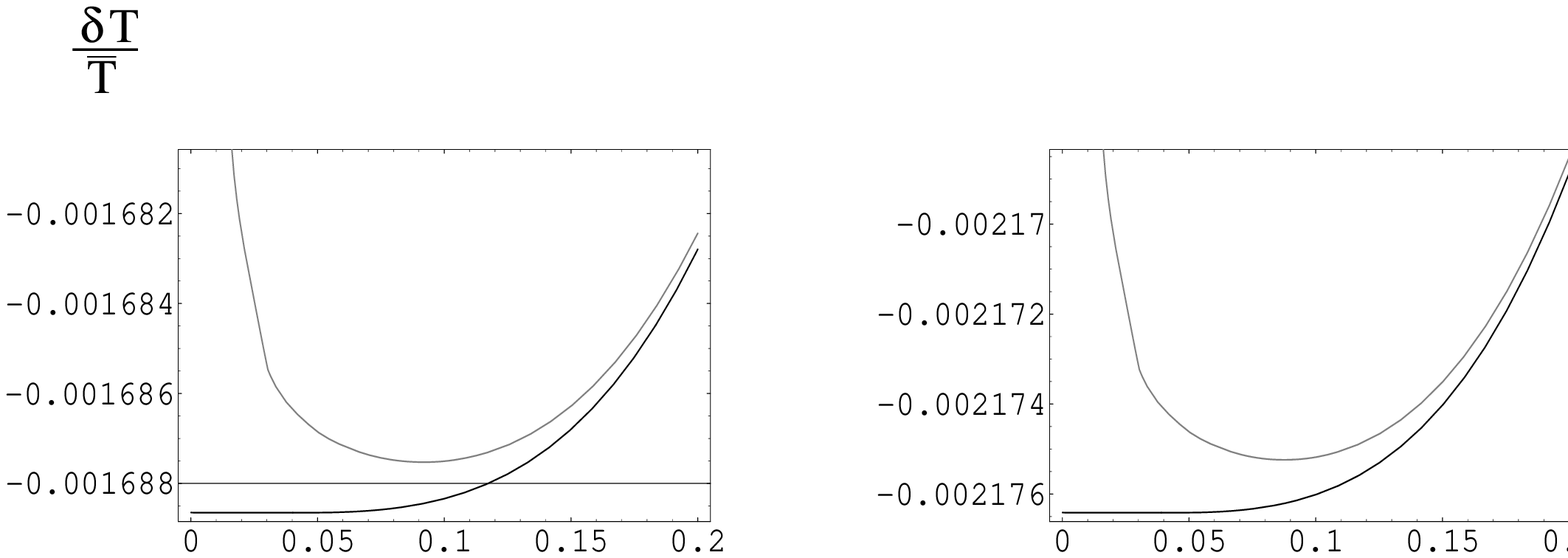}
\end{center}
\caption{$\frac{\delta T}{\bar T}$ for $z=0$ as a function of $\sigma$ for varying 
lattice constants 
(black curve: grid with 1000 points; gray curve: grid with 100 points). 
The width of the initial Gaussian distribution is $w=10^{-2}$,
the initial redshift is $z_i=20$, and the 
left (right) panel corresponds to $k=0.01868\,\bar{T}_0^2/H_0^2$ ($k=0.01449\,\bar{T}_0^2/H_0^2$).
\label{Fig-9}}      
\end{figure}

In Figs.\,\ref{Fig-6} and \ref{Fig-7} we show $\frac{\delta T}{\bar T}$ as a function 
of $\sigma$ at $z=0$ and $z=1$ when varying the shape of the initial profile at $z_i=20$. 
Our results are practically independent of these initial conditions. Again, it is seen 
that the major contribution to the profile at $z=0$ is being 
built up for $0\le z\le 1$.   
\begin{figure}
\begin{center}
\leavevmode
\leavevmode
\vspace{4.5cm}
\includegraphics{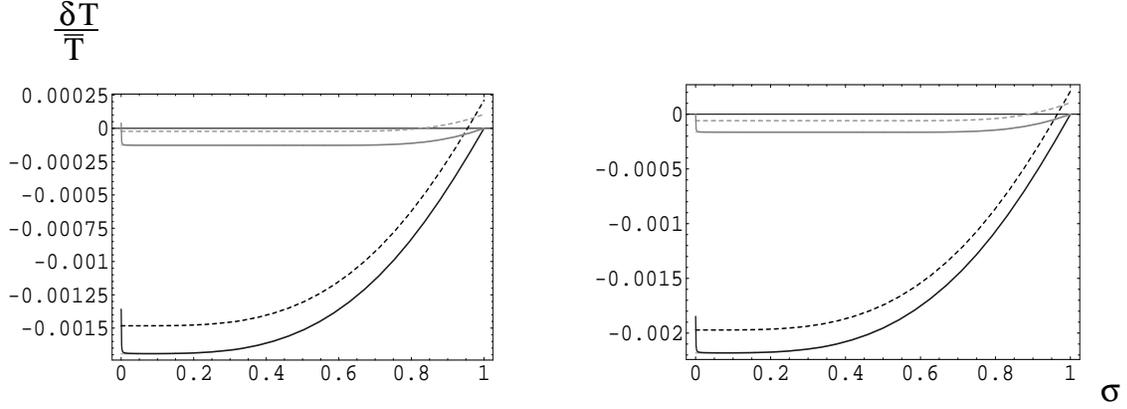}
\end{center}
\caption{$\frac{\delta T}{\bar T}$ for two values of $z$ as a function of $\sigma$:  
$z=0$ (black curves: solid line contains the 
cases $w=10^{-1}$ and $w=10^{-4}$, there is practically no difference; dashed curve corresponds to 
$w=\infty$) and $z=1$ (gray curves: same as for black curves). The initial 
redshift is $z_i=20$, the left (right) panel corresponds 
to $k=0.01868\,\bar{T}_0^2/H_0^2$ ($k=0.01449\,\bar{T}_0^2/H_0^2$).
\label{Fig-6}}      
\end{figure}
\begin{figure}
\begin{center}
\leavevmode
\leavevmode
\vspace{4.5cm}
\includegraphics{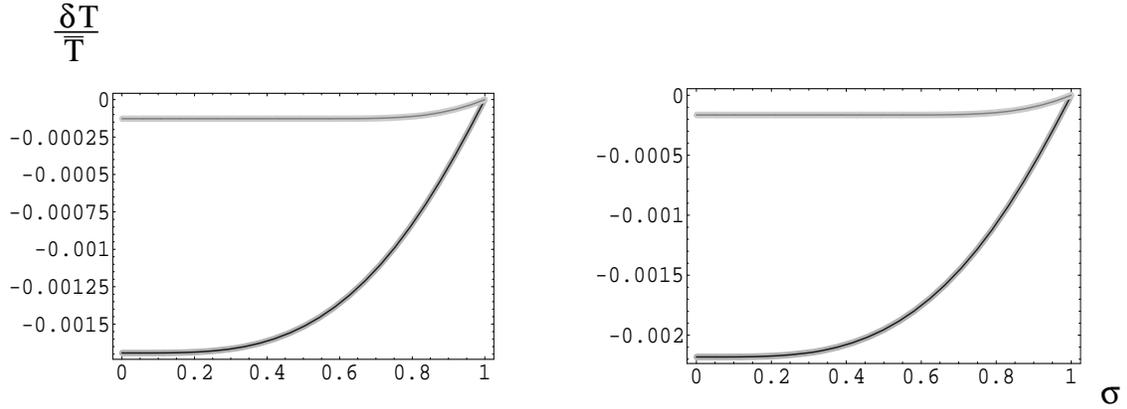}
\end{center}
\caption{$\frac{\delta T}{\bar T}$ for two values of $z$ as a function of $\sigma$:  
$z=0$ (two lower curves: initial Gaussian distribution 
with width $w=10^{-2}$ (black) and vanishing initial distribution (gray); 
$z=1$ (two upper curves: initial Gaussian distribution 
with width $w=10^{-2}$ (dark gray) and vanishing initial distribution (light gray);). 
The initial redshift is $z_i=20$, the left (right) panel 
corresponds to $k=0.01868\,\bar{T}_0^2/H_0^2$ ($k=0.01449\,\bar{T}_0^2/H_0^2$).
\label{Fig-7}}      
\end{figure}
Next, we investigate the dependence of the distribution on changes in $z_i$. 
Fig.\,\ref{Fig-8} shows $\frac{\delta T}{\bar T}$ as a function 
of $\sigma$ at $z=0$ and $z=1$ for \(z_i=5\), \(z_i=20\), and \(z_i=40\). 
Obviously, there is hardly any dependence on $z_i$. 
The plots in Fig.\,\ref{Fig-6} indicate a discontinuity at \(\sigma=0\). 
As demonstrated by Fig.\,\ref{Fig-9}, this is an artefact of the 
finite lattice constant when 
solving Eq.\,(\ref{difeqO}).

\section{Conclusion and Outlook\label{conclusion}}

In the present article we have discussed a model for the 
temperature-temperature correlation in the cosmic 
microwave background (CMB) at large angular 
separation. The key idea is to relate the 
discrepancy between the observed and the 
CMB-inferred velocity of the Local Group 
to a dynamic component in the 
CMB dipole. 

The origin of this dynamic component is tied to an 
anomaly in black-body spectra as 
it is predicted by deconfining 
SU(2) Yang-Mills thermodynamics when 
postulating that $SU(2)_{\tiny\mbox{CMB}}\stackrel{\tiny\mbox{today}}=U(1)_Y$, see 
\cite{Hofmann20051,Hofmann20052,GiacosaHofmann2005,SHG20061,SHG20062}.
The black-body anomaly, in turn, is computed in the effective 
theory for deconfining 
SU(2) Yang-Mills thermodynamics \cite{Hofmann20051} in terms of a one-loop diagram 
fixing the (on-shell) polarization tensor of the massless mode 
\cite{SHG20062}. As it seems, the 
polarization tensor for the massless mode is one-loop 
exact \cite{KH2007}. What is described by the polarization 
tensor in the effective theory is, on the microscopic 
level, the effect on the thermal spectral 
intensity of photon scattering off 
of electrically charged monopoles. The rate of change of the modified as compared to the
conventional black-body 
spectrum in dependence of mean temperature $\bar{T}$ is maximal at
$\bar{T}\sim 2\times T_c\sim 2\times 2,73\,$K or redshift 
$z\sim 1$. Primordial temperature inhomogeneities 
are enforced or weakened on the scale of $\frac{\delta T}{\bar{T}}\sim
10^{-2}$ in this regime leading to the emergence of a spherical profile
maximal at a center and vanishing for distances larger than 
horizon scale away from this center. The radial gradient of this 
profile is interpreted as a dynamic contribution to the dipole in the 
CMB temperature anisotropy thus accomodating the discrepancy 
between the directly observed and the inferred (by virtue of the realtivistic
Doppler effect) velocity of the the Local Group w.r.t. the CMB
rest frame. The rapid built-up of the spherical profile would then be
responsible for `inflating away' 
primordial large-scale anisotropies thus explaining the the 
missing power observed by WMAP \cite{WMAP3}: For a 2D analogue 
imagine a fluctuating rubber scarf enframed by a spherical 
boundary. Large-scale fluctuations are smoothened out by the process of 
pinching the scarf centrally and quickly pulling it out of 
the plane into a conical profile. In addition, the resulting and suppressed large-scale 
fluctuations would necessarily be correlated due to the common 
cause for their suppression \cite{Copi2007}.  

Our numerical simulations of the spherically symmetric case indicate
that the results are very robust against changes in the initial 
conditions. The main contribution to the built-up of the temperature 
profile arises for $z\sim 1$. Assuming an adiabatically slow cooling of
the CMB driven by a $\Lambda$CDM background cosmology, 
this corresponds to the regime in temperature $\bar{T}$ where 
$\frac{\delta T}{\bar{T}}$ changes most rapidly, see Fig.\,\ref{Fig-Intro}.  

 To substantiate the here-developed scenario further a dedicated simulation
of Eq.\,(\ref{eqmotion}) will be needed. Our future work thus 
will focus on the computation of dipole-subtracted 
large-angle correlations based on the model proposed here.    

\section*{Acknowledgments}
We would like to thank Francesco Giacosa, Frans Klinkhamer, Markus Schwarz, 
and Eduard Thommes for useful conversations. Helpful comments on the 
manuscript by Markus Schwarz are gratefully acknowledged.


\begin{thebibliography}{9}

\bibitem{PVLAS}
E. Zavattini {\sl et al.}, Phys. Rev. Lett. {\bf 96}, 110406 (2006). 

\bibitem{WMAP3}
A. Kogut et al., Astrophys. J. Suppl. {\bf 148}, 161 (2003).\\ 
D. N. Spergel et al., Astrophys. J. Suppl. {\bf 148}, 175 (2003).\\ 
D. N. Spergel et al., astro-ph/0603449.\\ 
L. Page et al., astro-ph/0603450.\\ 
G. Hinshaw et al., astro-ph/0603451.\\ 
N. Jarosik et al., astro-ph/0603452.  
 
\bibitem{knee}
L. B. G. Knee and C. M. Brunt, Nature {\bf 412}, 308 (2001).

\bibitem{Hofmann20051}
R. Hofmann, Int. J. Mod. Phys. A{\bf 20}, 4123 (2005), Erratum-ibid A{\bf 21}, 6515 (2006).\\  
R. Hofmann, Mod. Phys. Lett. A{\bf 21}, 999 (2006), Erratum-ibid. A {\bf 21}, 3049 (2006).

\bibitem{Hofmann20052}
R. Hofmann, PoS {\bf JHW2005}, 021 (2006) [hep-ph/0508176].

\bibitem{GiacosaHofmann2005}
F. Giacosa and R. Hofmann, Eur. Phys. J. C {\bf 50}, 635 (2007). 

\bibitem{Diakonov2004} 
D. Diakonov, N. Gromov, V. Petrov, S. Slizovskiy, Phys. Rev. D {\bf 70},
036003 (2004).

\bibitem{SHG20061}
M. Schwarz, R. Hofmann, and F. Giacosa, JHEP {\bf 0702}, 091 (2007).

\bibitem{SHG20062}
M. Schwarz, R. Hofmann, and F. Giacosa, Int. J. Mod. Phys. A{\bf 22},
1213 (2007). 

\bibitem{KH2007}
D. Kaviani and R. Hofmann, Mod. Phys. Lett. A {\bf 22}, 2343 (2007).

\bibitem{Hofmann2006}
R. Hofmann, hep-th/0609033.

\bibitem{SHGS2007}
M. Szopa, R. Hofmann, F. Giacosa, and M. Schwarz, arXiv:0707.3020
[hep-ph]

\bibitem{KortalsAltes}
C. P. Korthals Altes (Marseille, CPT), 
in *Minneapolis 2006, Continuous advances in QCD* 266-272
[hep-ph/0607154].\\ 
C. Korthals-Altes and A. Kovner, Phys. Rev. D {\bf 62}, 096008 (2000)
[hep-ph/0004052].\\ 
C. Korthals-Altes, hep-ph/0406138.

\bibitem{GH2007}
F. Giacosa and R. Hofmann, Phys. Rev. D {\bf 76}, 085022 (2007).

\bibitem{Hofmann2007M}
R. Hofmann, arXiv:0710.0962 [hep-th].

\bibitem{Riess1998} 
A. G. Riess et al., Astron. J. \textbf{116}, 1009 (1998).

\bibitem{Perlmutter1998} 
S. Perlmutter et al., Astrophys. J. \textbf{483},
565 (1998).

\bibitem{Schmidt1998} 
B. P. Schmidt et al., Astrophys. J. \textbf{507}, 46
(1998).

\bibitem{Spergel}
D. N. Spergel et al., astro-ph/0603449.

\bibitem{Atiyah}
M. F. Atiyah and N. J. Hitchin, Phil. Trans. Roy. Soc. Lond. A{\bf 315}, 459 (1985).\\ 
M. F. Atiyah and N. J. Hitchin, Phys. Lett. A{\bf 107}, 21 (1985).

\bibitem{GiacosaHofmann2007}
F. Giacosa and R. Hofmann, hep-th/0703127.

\bibitem{2MRS}
P. Erdogdu et al., Mon. Not. Roy. Astron. Soc.{\bf 373}, 45 (2006) [astro-ph/0610005].\\ 
P. Erdogdu et al., talk given at 41st Rencontres de Moriond, 
Workshop on Cosmology: Contents and Structures of the Universe, La Thuile, Italy, 18-25 Mar 2006 [astro-ph/0605343].
\bibitem{Peebles}
P. J. Peebles and D. T. Wilkinson, Phys. Rev.{\bf 17}, 2168 (1968).

\bibitem{Hinshaw}
G. Hinshaw et al., astro-ph/0603451.

\bibitem{PDG2006}
Review of Particle Physics, Particle Data Group, p. 98 (2006).

\bibitem{Kogut93}
A. Kogut et al., Astrophys. J. \textbf{419}, 1 (1993).

\bibitem{Schiesser}
W. E. Schiesser, Computational mathematics in Engineering and Applied Science: ODEs, DAEs and PDEs, 
CRC Press. (1994). 

\bibitem{Tkatchev}
S. Yu. Khlebnikov and I. I. Tkachev, Phys. Rev. Lett. {\bf 77}, 219 (1996). \\ 
S. Khlebnikov and I. I. Tkachev, Phys. Rev. Lett. {\bf 79}, 1607 (1997). 

\bibitem{Prokopec}
T. Prokopec and T. G. Roos, Phys. Rev. D{\bf 55}, 3768 (1997).

\bibitem{Copi2007}
C. J. Copi, D. Huterer, D. J. Schwarz, and G. D. Starkman, Phys. Rev. D{\bf 75}, 023507 (2007). 

\end{thebibliography}
\end{document}